# Role of Polarons in Single-Atom Catalysts: Case Study of $Me_1$ [$Au_1$, $Pt_1$, and $Rh_1$] on $TiO_2(110)$

Panukorn Sombut[1], Lena Puntscher[1], Marlene Atzmueller[1], Zdenek Jakub[1], Michele Reticcioli[2], Matthias Meier[1,2], Gareth S. Parkinson[1] and Cesare Franchini[2,3*]

[1]Institute of Applied Physics, TU Wien, 1040 Vienna, Austria

[2]Faculty of Physics, Center for Computational Materials Science, University of Vienna, 1090 Vienna, Austria

[3]Alma Mater Studiorum, Università di Bologna, 40127 Bologna, Italy

* E-mail: cesare.franchini@univie.ac.at

**Abstract**

The local environment of metal-oxide supported single-atom catalysts plays a decisive role in the surface reactivity and related catalytic properties. The study of such systems is complicated by the presence of point defects on the surface, which are often associated with the localization of excess charge in the form of polarons. This can affect the stability, the electronic configuration, and the local geometry of the adsorbed adatoms. In this work, through the use of density functional theory and surface-sensitive experiments, we study the adsorption of $Rh_1$, $Pt_1$, and $Au_1$ metals on the reduced $TiO_2(110)$ surface; a prototypical polaronic material. A systematic analysis of the adsorption configurations and oxidation states of the adsorbed metals reveals different types of couplings between adsorbates and polarons. As confirmed by scanning tunneling microscopy measurements, the favored $Pt_1$ and $Au_1$ adsorption at oxygen vacancy sites is associated with a strong electronic charge transfer from polaronic states to adatom orbitals, which results in a reduction of the adsorbed metal. In contrast, the $Rh_1$ adatoms interact weakly with the excess charge, which leaves the polarons largely unaffected. Our results show that an accurate understanding of the properties of single-atom catalysts on oxide surfaces requires a careful account of the interplay between adatoms, vacancy sites, and polarons.

1. Introduction

Due to their particular local environment, single-atom catalysts (SACs) represent a new frontier in heterogeneous catalysis, resulting in a unique electronic structure in comparison with supported nanoparticle catalysts [1–7]. Metal atoms adsorbed on solid supports and their resulting catalytic properties combine the advantages of homogeneous catalysts (high activity and selectivity) and heterogeneous catalysts (stable and easy to separate), while minimizing the amount of precious metal used in heterogeneous catalysis [3, 8]. Therefore, SACs are expected to bridge the gap between heterogeneous and homogeneous catalysts. However, the tendency of isolated atoms to aggregate into small clusters due to their relatively high surface energy is problematic. A strong covalent metal-support interaction is capable of stabilizing SACs [9]. However, adsorption of SACs in atomic defects on the substrate surface is the most effective way to avoid clustering and stabilize isolated metal atoms on the support [10, 11], expanding the applicability and efficiency of single-atom catalysis.

Nonetheless, such defective surfaces can affect the properties of adsorbed adatoms, whether within or outside the defect itself and must be carefully investigated. Further, the presence of point defects on transition-metal oxide surfaces can inject excess electrons which can locally couple with ionic vibrations and form small polarons [12]. Adsorbate/oxide-surface interactions are known to be significantly affected by defects and their associated polarons [13–15], but their effect on catalysis has been rarely considered [16, 17]. For instance, the properties of metal atom ($Me_1$) species on reduced rutile $TiO_2(110)$ surface, a prototypical polaronic system, has been extensively studied as a SAC [18–23], but the potential effects of polarons are generally neglected. On the reduced $TiO_2(110)$ surface, polarons tend to localize at a 6-fold coordinated Ti atom ($Ti_{6c}$) in the subsurface layer in the vicinity of the 2-fold coordinated oxygen vacancy ($V_{O2c}$) site, reducing $Ti^{4+}$ to $Ti^{3+}$ ions [24]. At elevated temperatures, polaron diffusion between subsurface and surface layers may occur, altering the properties and nature of the polaronic state [24–26], and potentially affecting the stability and properties of the adatoms as well.

In recent years, computational studies have become a powerful tool to accurately describe catalytic reactions at the atomic scale in heterogeneous catalysis [27–29]. In particular, first-principles calculations, within the density functional theory (DFT) framework, have revealed several useful insights into the nature of active sites and the reaction mechanisms in the SAC models [30]. Furthermore, the advances brought about by DFT studies facilitate the interpretation of experimental measurements, and might propose specific substrate materials and metal atoms as optimal candidates for efficient SAC processes.

The purpose of this study is to investigate the effect of polarons on the stability and properties of single-metal atom catalysts. We consider the adsorption of $Rh_1$, $Pt_1$, and $Au_1$ transition metals on the reduced rutile $TiO_2(110)$ surface. To investigate the interplay between electron polarons, oxygen vacancies, and the adatom on the $TiO_2(110)$ surface, we performed DFT+U calculations and compared the results with experimental data taken from existing literature ($Au_1$ [31–35]) as well as new scanning tunneling microscopy data for the $Pt_1$ and $Rh_1$ systems. Accordingly, we confirm that charge transfer occurs for $Pt_1$ and $Au_1$ adatoms located in O vacancies ($V_O$), making them the preferred adsorption configurations, by carefully studying the most stable adsorption sites, the appropriate oxidation states, and the interactions among adatoms, O vacancies, and polarons. A low diffusion barrier on the surface of $TiO_2(110)$ allows $Pt_1$ adatoms to reach O vacancies when dosed in low amount at room temperature. $Au_1$ adatoms exhibit the same behavior but at a lower temperature with an even lower diffusion barrier. $Rh_1$ is found to have no preference for such defects, leaving the polarons essentially unaffected in the subsurface. Our results show that the properties of single-atom catalysts on metal-oxide surfaces can be accurately described only by carefully considering the interaction with point defects and polarons, as well as the reduction of the adsorbed metals.

2. Methods

2.1 Computational methods

All calculations were performed by using the Vienna *ab initio* simulation package (VASP) [36, 37]. The projector augmented wave method [38, 39] was used for the electron and ion interaction, with the plane-wave basis set cutoff energy set to 400 eV, optimized to include van der Waals interactions as proposed by Dion *et al* [40] with the optimized functional (optPBE-DF) [41, 42]. However, DFT calculations have known drawbacks when dealing with electron localization effects [43]. Therefore, it is preferable to use first-principle schemes that account for the localized charge, such as the DFT+U method used here within [44, 45]: we dressed the d orbitals of the Ti atoms with an effective on-site coulomb repulsion term ($U_{eff}$ of 3.9 eV) [46], previously determined by constrained-random-phase-approximation calculations in bulk rutile [25]. The unreconstructed rutile surface was modelled using an asymmetric slab containing five $TiO_2$ tri-layers in a large two-dimensional 6×2 unit cell and including a vacuum space region greater than 12 Å along the z-axis. The top three tri-layers were allowed to relax, while the bottom two tri-layers were kept fixed at their bulk positions. An alternative slab model in which the broken bonds at the bottom layer were saturated by pseudo-hydrogen atoms did not affect our conclusions regarding the adsorption energies and polaron stabilities. The convergence is achieved when the electronic energy step of $10^{-5}$ eV is obtained and forces acting on ions become smaller than 0.01 eV/ Å. The adsorption energies were computed according to the formula:

$$E_{ads} = E_{TiO_2(110)+adatom} - (E_{TiO_2(110)} + E_{adatom})$$

Where $E_{TiO_2(110)+adatom}$ is the total energy of the $TiO_2(110)$ surface with the adsorbed adatom, $E_{TiO_2(110)}$ is the total energy of the clean $TiO_2(110)$ surface with an oxygen vacancy and the most stable polaronic configuration [47, 48]. The $E_{adatom}$ represents the energy of the atom in the gas phase. The surface slab is displayed in Fig. 1a.

The diffusion barriers of an adatom on the reduced $TiO_2(110)$ surface were evaluated using the climbing image nudged elastic band (CI-NEB) method [49, 50] with three interpolated images. As initial and end states we carefully selected solutions including adatoms in the same oxidation state.

In order to inspect the different charge states of the metal atoms on different adsorption sites and the effect of the presence of polarons, we adopted the following strategy.

1.) We altered the charged-neutral state of the reduced $TiO_2(110)$ surface in order to fully saturate the excess electrons donated by the $V_O$. For instance, two holes were added to neutralize two excess electrons donated by the $V_O$. This setup allows us to study the adsorbed metal atoms with different valence states on the reduced surface without considering the perturbation due to polaron charge trapping.

2.) Adsorption of $Me_1$ at the known adsorption sites, according to literature [22, 23, 31, 33–35, 51–54], was tested with different oxidation states. By artificially changing the total amount of electrons, we changed the valence of the adatom. For instance, the adsorbed $Me^0$, $Me^-$, and $Me^{2-}$ were obtained by adding zero, one, and two extra electrons, respectively. All possible electronic and coordination configurations were modeled via geometry relaxation, and only the stable local minima were retained. This allows us to obtain the preferred oxidation state of the adsorbed adatom at the specific adsorption sites.

3.) The retained structures from the 2$^{nd}$ step have their charge renormalized to match the charge of the respective system with one oxygen vacancy and two excess electrons. To selectively control the negative charge localization, we used the occupation matrix control tool [55]. This method allows to control the charge localization by specifying the $Ti^{3+}$ d orbital occupation. For example, $Pt^0$ or $Pt^-$ adsorbed at $V_O$ with two or one additional electrons to form polarons, respectively. Alternatively, $Pt^{2-}$ at $V_O$ with no additional electron. As a first step, the occupation matrix control was applied. Next, the constraint was removed: the structures were recalculated starting from the previous wavefunction. We carefully checked whether the final unconstrained calculations led to the expected solution for the valence charge (in same case, we were not able to obtain all possible valence states for the metal atoms, as described case by case in the Results section). For example $Pt_1$ at $V_{O2c}$;

(i) Same solutions: $Pt^0$ with two polarons as an initial configuration leads to $Pt^{2-}$ without polarons after removing the constraint and relaxation. $Pt^{2-}$ without polarons as initial configuration remains $Pt^{2-}$ after optimization. $Pt_1$ manifests a clear preference for $Pt^{2-}$ as oxidation state when located at $V_{O2c}$.

(ii) Different solutions: $Pt^-$ with only one polaron as an initial configuration stays as $Pt^-$ after removing the constraint and optimization, at the same site as previously mentioned. In this case, we compare the adsorption energy of $Pt^-$ and $Pt^{2-}$ and retain the $Pt^{2-}$ at $V_{O2c}$ as the most stable oxidation state for the next step.

4.) Using the three best configurations of each metal adatom from the 3$^{rd}$ step, a set of new calculation is conducted to include and study the effect of polarons on adatoms. In this case, polaron positions are being tested with respect to the adatom location. The adatom adsorption energy is affected only by polarons in layers S0 and S1, but different types of polarons at various Ti sites in these layers exist. The type or orbital character of polarons depends on the orbital topology, degree of localization, and associated local structural distortions [24].

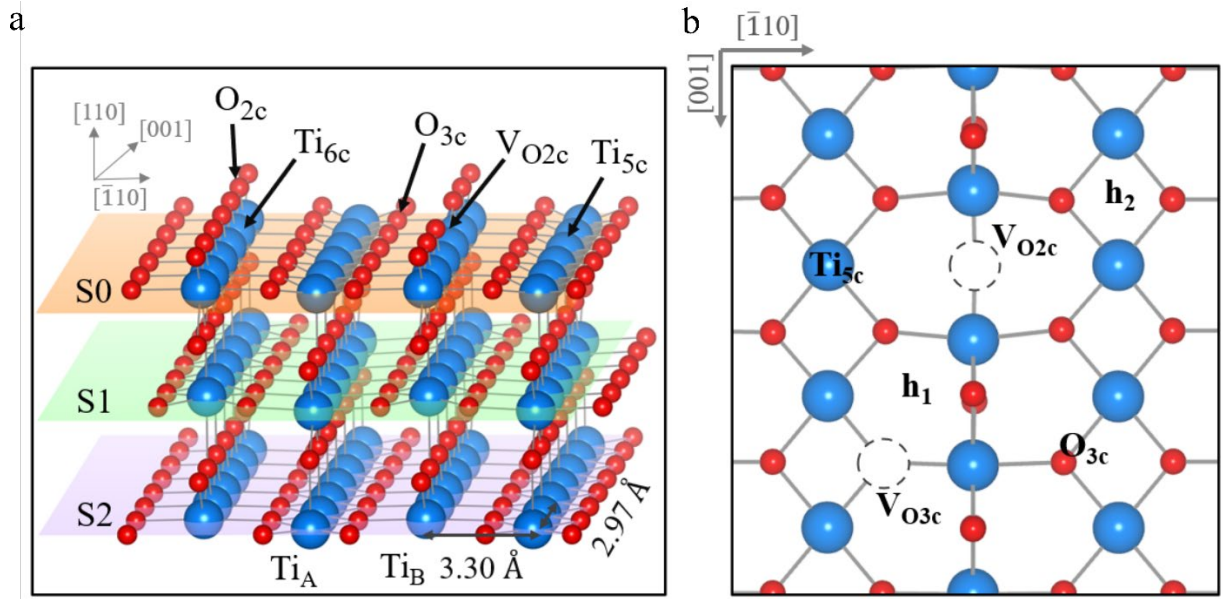

**Fig. 1** Structure of the $TiO_2(110)$ surface, Ti atom, O atom, and oxygen vacancy are depicted in big blue sphere, small red sphere and dashed-circle, respectively. a) side view of a 6×2 unit cell and b) top view with possible adsorption sites: a 2-fold oxygen vacancy ($V_{O2c}$), hollow$_1$ ($h_1$), hollow$_2$ ($h_2$), a 5-fold titanium atom ($Ti_{5c}$), a 3-fold oxygen atom ($O_{3c}$) and a 3-fold oxygen vacancy ($V_{O3c}$).

### 2.2 Experimental details

Low temperature STM was performed in a two-vessel UHV chamber consisting of a preparation chamber (p<1×10$^{−10}$ mbar) and an analysis chamber (p<2×10$^{−11}$ mbar). The preparation chamber is equipped with a commercial

XPS. The analysis chamber is equipped with an Omicron LT-STM with a Qplus sensor and an in-vacuum preamplifier [56]. Room-temperature STM was performed in a second two-vessel UHV chamber consisting of a preparation chamber (p<1×10$^{-10}$ mbar) and an analysis chamber (p<5×10$^{-11}$ mbar). The analysis chamber is equipped with a nonmonochromatic Al Kα X-ray source (VG), a SPECS Phoibos 100 analyzer for XPS, and an Omicron μ-STM. STM in both UHV chambers was conducted in constant current mode with an electrochemically etched W tip on synthetic TiO$_2$(110) single crystals (from CrysTec GmbH) prepared in UHV by sputtering (1 kV, Ar+, 15 min) and annealing (20 min, 700 °C). Rh and Pt were deposited using an e-beam evaporator (FOCUS), with the flux calibrated using a temperature-stabilized quartz microbalance (QCM). The STM images were corrected of distortion and creep of the piezo scanner as described in ref [57].

3. Results

The rutile TiO$_2$(110) surface is one of the most intensively studied metal-oxide surfaces [58–60]. This surface layer consists of bridging oxygen rows (O$_{2C}$) and 5-fold coordinated titanium (Ti$_{5c}$) rows (see the structural model in Fig. 1). Bridging oxygen vacancies (V$_{O2c}$) can be easily created in UHV conditions by sputtering and annealing. Each V$_{O2c}$ defect donates two excess electrons to the surface, which are trapped in Ti sites forming small polarons, clearly identified by shape in-gap peaks [12, 48, 61–63]. We aim to elucidate the impact of polarons on the stability and properties of single-metal atoms (Pt$_1$, Au$_1$, and Rh$_1$) adsorbed on the rutile TiO$_2$(110) surface. The adsorption sites are labeled in Fig. 1b.

3.1 Pt$_1$ on the TiO$_2$(110) surface

In order to investigate the adsorption of Pt$_1$ on TiO$_2$(110), we considered possible adsorption sites as reported in the available literature [21, 22, 51, 52] but including the presence of polarons. Figure 2a-c shows the three most stable adsorption sites and their corresponding calculated projected DOS (pDOS) as well as their oxidation states, taking polaronic effects into account. Pt$_1$ adsorbed at a bridging oxygen vacancy is the most stable adsorption site (E$_{ads}$ = −3.22 eV). The Pt$_1$ atom in this configuration shows an oxidation state of Pt$^{2-}$ due to the charge transfer of two excess electrons from the reduced surface to the Pt$_1$ adatom, meaning that it is more favorable to transfer the electron to Pt$_1$ rather than using the excess charge to form polarons. The calculated projected DOS indeed shows no in-gap Ti polaronic peaks. We note that Pt$^-$ in this configuration is less stable than Pt$^{2-}$ by 0.39 eV (Fig. S6). Pt$_1$ at the hollow(1) site is next in energy (E$_{ads}$ = −2.83 eV), with Pt$^0$ being the most stable oxidation state. In this case, the two excess electrons prefer to be trapped in polaronic sites, and no evident net charge transfer to Pt$_1$ occurs. The projected DOS also shows two polaronic in-gap peaks for two Ti$^{3+}$ sites, similar to the clean surface (Fig. S1). When polaron formation is maintained, as in the case of Pt$^0$, polarons prefer to be trapped near the oxygen vacancy in the S1 layer [24]. The last possible adsorption site with comparatively large adsorption energy is Pt$_1$ adsorbed atop a 3-fold coordinated oxygen atom on the basal plane. Pt$^-$ is the most stable oxidation state at this site, with one polaron again preferably located in the S1 layer near the oxygen vacancy. Overall, the polarons prefer to be located in the S1 layer in the proximity of the V$_{O2c}$ and maximize their distance with the adsorbed Pt$_1$ adatom due to the repulsive interaction between the negatively charged polaron and the adsorbed metal atom. A variation of the order of 300 meV in the adatom adsorption energy can be seen depending on polaron position and its distance to the adatom (Fig. S2).

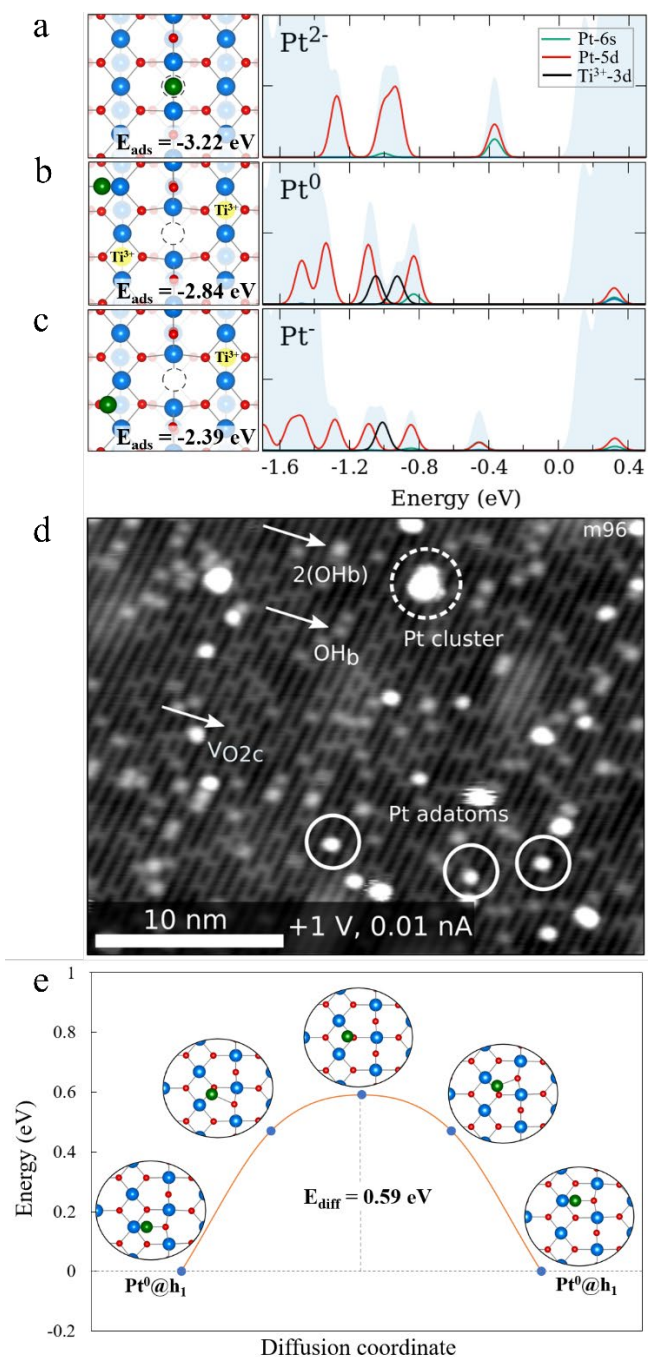

**Fig. 2** Minimum energy configuration of possible adsorption sites of $Pt_1$ on the $TiO_2(110)$ surface. **a)** $Pt_1@V_{O2c}$ without polarons resulting in a $Pt^{2-}$ configuration. **b)** $Pt_1$ in $h_1$ with two polarons resulting in a $Pt^0$ configuration. **c)** $Pt_1$ atop $O_{3C}$ atom with one polaron resulting in a $Pt^-$ configuration, where O, $Ti^{4+}$, $Ti^{3+}$(polaron), $Pt_1$ are small red, big blue, big yellow, big green spheres and $V_O$ is a dashed-circle, respectively. Each configuration is aligned with its respective DOS panels, where the total DOS, pDOS of Pt (5d), pDOS of Pt (6s), and pDOS of $Ti^{3+}$(3d, polarons) are filled light blue, red line, green line, and black line, respectively. **d)** Empty-state RT STM images of 0.007 ML Pt on the reduced rutile $TiO_2(110)$ surface deposited at room temperature, with surface oxygen vacancies ($V_{O2c}$), OH groups ($OH_b$), pairs

of OH groups (2(OH$_b$)), Pt adatoms (circle) and Pt clusters (dashed circle) labelled in the image. And **e)** diffusion path of Pt$_1$ adatom on TiO$_2$(110) from one hollow site to the next in the neighboring unit cell without the perturbation from any V$_O$.

The room-temperature STM image (Fig. 2d) shows Pt$_1$ adsorbed on the reduced rutile TiO$_2$(110) surface. The TiO$_2$(110) surface is characterized in STM by bright rows of 5-fold coordinated Ti$^{4+}$ alternating with dark rows of two-fold bridging O$^{2-}$, which run along the [001] direction [60]. Small protrusions over the dark rows can be assigned to oxygen vacancies V$_{O2c}$, and brighter features on the dark rows can be assigned to bridging OH [64] and pairs of bridging OH. These originate from water dissociation at the V$_{O2c}$ [65–67]. Pt adatoms can be stabilized at low coverage, but clusters dominate the higher the coverage. The Pt adatoms adsorb in the 2-fold oxygen vacancies (marked in circle), which can be identified directly when rare events of adatoms diffusion occur. In this case, the Pt adatom changes from one V$_O$ to another (Fig. S7). Both the initial and final vacancy are imaged in STM. Dosing water or oxygen gas (O$_2$) at room temperature (2 L; 100 s 2×10$^{-8}$ mbar) prior to the Pt deposition leads to the reparation of the V$_O$ site [66–68]. In such an experiment, only Pt clusters are observed (as seen in Fig. S8). These results show the direct correlation between oxygen vacancies and the Pt adatoms stabilization.

In order to account for dynamical effects involving diffusion of the metal species, we calculated the diffusion barrier of the adsorbed Pt$_1$ along the [001] direction. The calculated energy barrier is low (0.59 eV, see Fig. 2e) so that Pt$_1$ can diffuse already at room temperature and reach the best adsorption site (at the oxygen vacancies). The calculated results are therefore in-line with our room-temperature STM images, where only Pt$_1$ at oxygen vacancies have been observed on the reduced TiO$_2$(110) surface at low coverage. At lower dosing temperature, possible metastable adsorption configurations outside the V$_{O2c}$ can exist.

### 3.2 Au$_1$ on the TiO$_2$(110) surface

Au$_1$ adsorbed on TiO$_2$(110) exhibits a similar structure than Pt$_1$, as shown in Fig. 3. The most stable adsorption site for Au$_1$ is located at V$_{O2c}$ (E$_{ads}$ = −2.06 eV). Au$_1$ located at this site becomes negatively charged (Fig. 3a) with an oxidation state of Au$^-$. Polaronic configurations are most favorable when the remaining polaron forms in the S1 layer close to the V$_{O2c}$ (Fig. 3a). We also considered the on-top 5-fold Ti atom adsorption site, as it was previously considered in other theoretical works [31, 32, 69]. The most stable valence state of Au$_1$ at this site is Au$^-$ with one polaron remaining in the S1 layer close to V$_{O2c}$, as shown in Fig. 3b (E$_{ads}$ = −1.46 eV). The calculated projected DOS in both cases shows that the valence d and s states are filled, with one characteristic in-gap polaronic peak. Similarly to Pt$_1$, the different polaronic configurations can modify the adsorption energy up to 300 meV (Fig. S3-4).

The diffusion barrier of Au$_1$ on the TiO$_2$(110) from the 5-fold Ti atom to the next 5-fold Ti atom along [001] direction is low (0.2 eV), and even lower is the corresponding barrier for diffusing from the 5-fold Ti atom to the oxygen vacancy which is almost barrierless, indicating a facile diffusion (Fig. 3c). Based on our results, we can conclude that at very low Au coverage, all Au$_1$ adatoms will be trapped in the oxygen vacancies on the surface. This result agrees with the room temperature STM measurement at low coverage [35] and with the observation of the nucleation of gold clusters at oxygen vacancy sites at high coverage [31, 32].

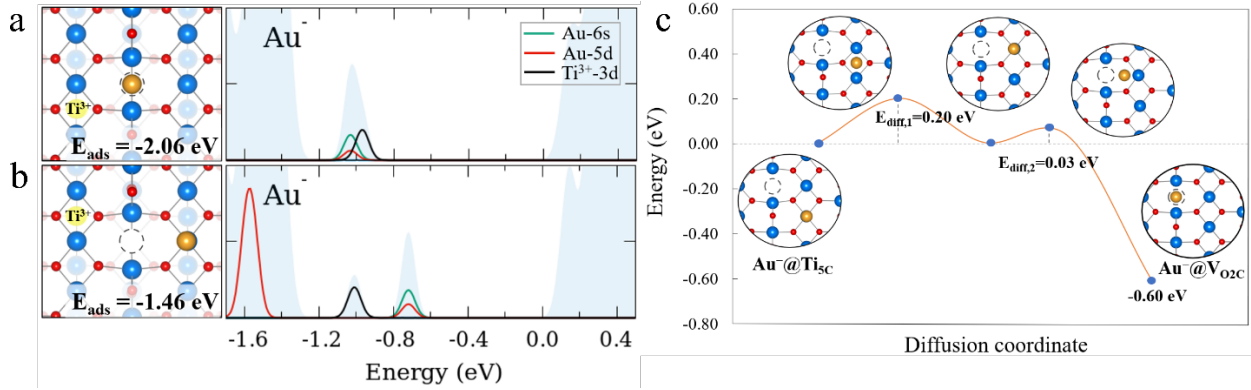

**Fig. 3** Minimum energy configuration of possible adsorption sites of $Au_1$ on the $TiO_2(110)$ surface **a)** $Au_1@V_{O2c}$ with one polaron resulting in a $Au^-$ configuration. **b)** $Au_1$ atop $Ti_{5c}$ atom with one polaron resulting in a $Au^-$ configuration. Each configuration is aligned with its respective DOS panels, where the total DOS, pDOS of Au (5d), pDOS of Au (6s), and pDOS of $Ti^{3+}$(3d, polarons) are filled light blue, red line, green line, and black line, respectively. And **c)** diffusion path of $Au_1$ atom on $TiO_2(110)$ from atop $Ti_{5c}$ to the $V_{O2c}$ site, where O, $Ti^{4+}$, $Ti^{3+}$(polaron) and $Au_1$ are small red, big blue, big yellow and big yellow-brown spheres, and $V_O$ is a dashed-circle, respectively.

### 3.3 $Rh_1$ on the $TiO_2(110)$ surface

We studied the adsorption of $Rh_1$ adatoms on $TiO_2(110)$ by combining DFT+U calculations with XPS and STM experiments. The adsorption sites with their respective adsorption energy are summarized in Fig. 4. Interestingly, $Rh_1$ adsorbed at $V_{O2c}$ (Fig. 4c) is not the preferential configuration ($E_{ads}= -2.82$ eV). We assign a $Rh^-$ state to this configuration, due to the presence of an in-gap polaronic peak from one $Ti^{3+}$ atom. Two configurations at hollow sites are shown in Fig. 4a and 4b. The oxidation state of the $Rh_1$ adatom at hollow sites depends on the nearest oxygen atoms binding to it. For instance, the relaxed structures show that $Rh^0$ binds to the $O_{3c}$ and $O_{2c}$ atoms ($E_{ads}= -3.24$ eV), whereas the $Rh^-$ at the hollow site near $V_{O2c}$ binds to only the $O_{3c}$ atom ($E_{ads} = -3.05$). A charge transfer occurs for $Rh_1$ near the $V_{O2c}$ where one excess electron transfers to $Rh_1$, leaving the second excess electron to form a polaron in the S1 layer near $V_{O2c}$. The projected DOS also shows that there is one characteristic in-gap polaronic peak. The $Rh_1$ at a hollow site that is distant from $V_{O2c}$ is assigned to $Rh^0$ and has two remaining polarons in a preferential S1 configuration. The calculated projected DOS shows the two in-gap polaronic peaks from $Ti^{3+}$ atoms. Polarons reside in the S1 layer minimizing the distance from $V_{O2c}$ while maximizing the distance from the $Rh_1$. However, the adsorption energies of $Rh^0$ and $Rh^-$ are almost degenerate.

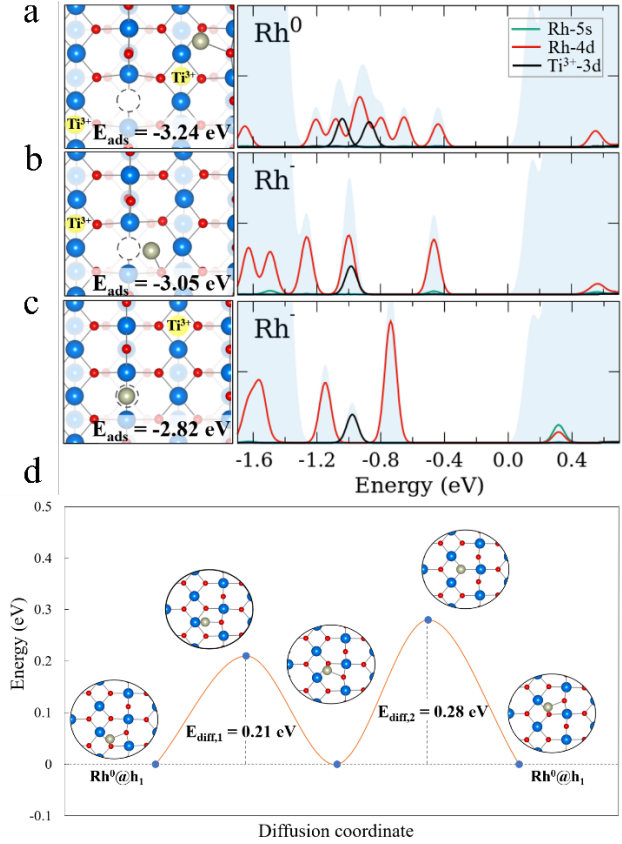

**Fig. 4** Minimum energy configuration of possible adsorption sites of $Rh_1$ on the $TiO_2(110)$ surface **a)** $Rh_1$ in $h_1$ with two polarons resulting in a $Rh^0$ configuration. **b)** $Rh_1$ in $h_1$ near the $V_{O2c}$ with one polaron resulting in a $Rh^-$ configuration. **c)** $Rh_1@V_{O2c}$ with one polaron resulting in a $Rh^-$ configuration. Each configuration is aligned with its respective DOS panels, where the total DOS, pDOS of Rh (4d), pDOS of Rh (5s), and pDOS of $Ti^{3+}$(3d, polarons) are filled light blue, red line, green line, and black line, respectively. And **d)** diffusion path of $Rh_1$ atom on $TiO_2(110)$ from the hollow(1) site to the next hollow(1) site, where O, $Ti^{4+}$, $Ti^{3+}$ and $Rh_1$ are small red, big blue, big yellow and big silver spheres, and $V_O$ is a dashed-circle, respectively.

The adsorption of an adatom at the 3-fold oxygen vacancy ($V_{O3c}$) has been proposed to rationalize the result from scanning transmission electron microscopy (STEM) experiments and DFT calculations for $Pt_1$ and $Rh_1$ [22, 23]. Other works argued that this adsorption does not exist on $TiO_2(110)$ due to the energetically unfavorable formation of the 3-fold oxygen vacancy ($V_{O3c}$) on the bare $TiO_2(110)$ surface [35, 52]. Our calculations also show that the formation of $V_{O2c}$ defects is much more favorable than the $V_{O3c}$ vacancy by 1.39 eV on the pristine surface. Figure 5 shows, however, that when $Rh_1$ is present and adsorbs at a $V_{O3c}$, it becomes the most thermodynamically stable adsorption site ($E_{ads} = -3.42$ eV, Fig. S5) with respect to the $V_{2Oc}$ formation.

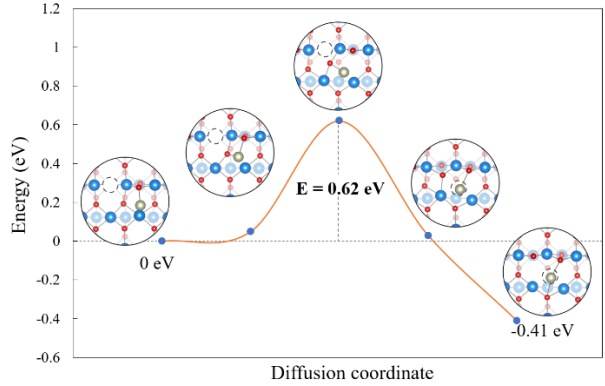

**Fig. 5** Energy profile of $V_O$ migration between the initial state ($Rh^-$ at hollow site) and the final state ($Rh^-$ at 3-fold $V_O$). In order to match the oxidation state of the final state in the NEB, the initial state is set to –1 instead of 0, giving an offset of ~0.2 eV.

Despite the unfavorable 3-fold oxygen vacancy formation energy found for the pristine surface, the presence of adatoms can alter the energetic cost of creating different vacancies other than $V_{O2c}$ and therefore should not be excluded. While $Rh_1$ in $V_{O3c}$ is the most stable configuration found, the question arises whether it can be reached or not, as the as-prepared surface does not exhibit such defects prior to the deposition of Rh. We calculated the oxygen migration from $V_{O2c}$ to $V_{O3c}$ with the presence of $Rh_1$ (Fig. 5), ignoring the presence of polarons and changes in the oxidation state of Rh. A barrier of 0.62 eV is obtained, significantly higher than the diffusion of Rh (0.28 eV, Fig. 4d) on the bare surface, suggesting that prior to the formation of in $V_{O3c}$ adsorbed Rh, the adatoms would sinter into clusters (assuming these would be favorable in energy).

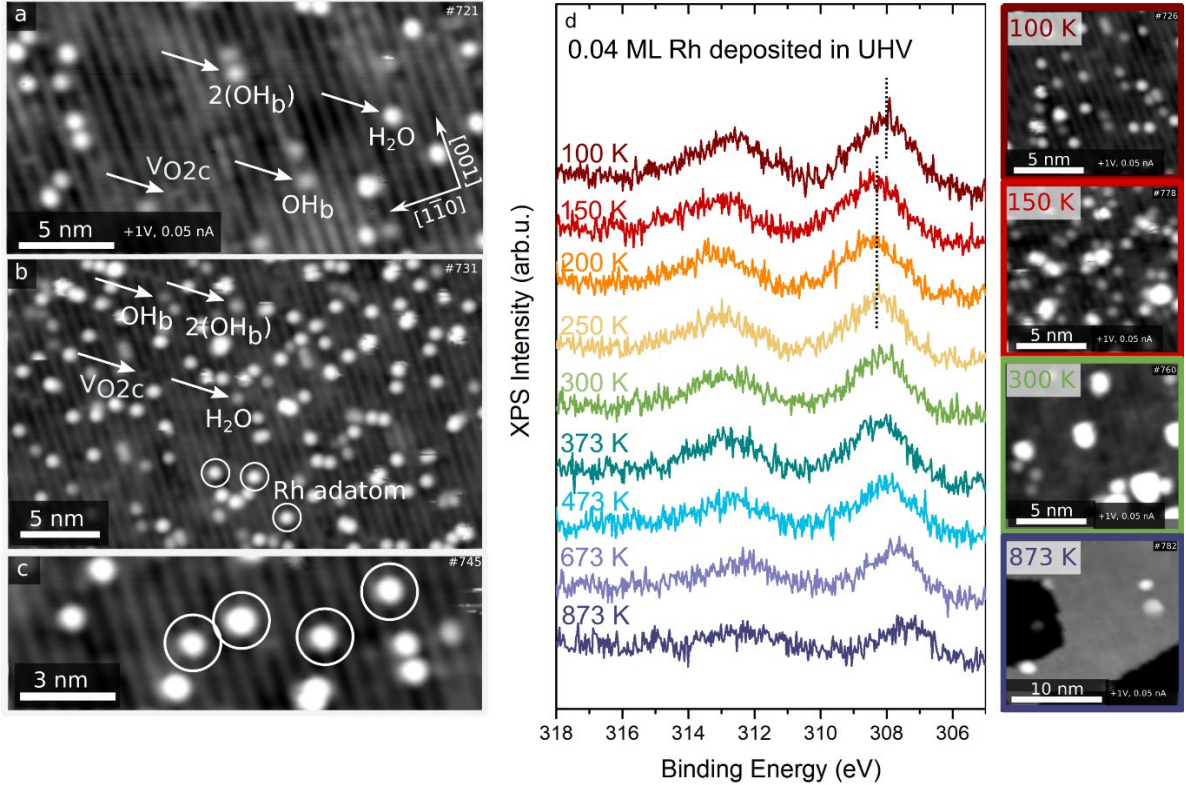

**Fig. 6** Experimental investigation of 0.04 ML Rh adsorbed on TiO$_2$(110). **a)** STM image (acquired at 78 K) of the clean TiO$_2$(110) surface. Surface oxygen vacancies (V$_{O2c}$), OH groups (OH$_b$), pairs of OH groups (2(OH$_b$)), and adsorbed water (H$_2$O) molecules can be seen. **b)** and **c)** Following the evaporation of 0.04 ML Rh at 100K, Rh adatoms (circle) can be seen. **d)** XPS spectra and STM images of the Rh 3d peak after gradually increasing the temperature.

Figure 6a-c shows STM images of TiO$_2$(110) before (a) and after deposition of 0.04 ML of Rh (b) and (c) at 100 K, where 1 ML corresponds to 1 Rh atom per surface unit cell. The images were acquired using liquid nitrogen as the cryogen for the LT-STM, giving a sample a temperature of 78 K. Features on the bright Ti$^{4+}$ rows (Fig 6a-b) with an apparent height of 90 – 100 pm can be assigned to molecular water adsorbed to Ti$^{4+}$ [70]. Rh adatoms (marked in circles) are located atop the bright Ti$^{4+}$ rows but slightly tilted towards the dark rows (bridging O$^{2-}$). All Rh adatoms adsorb at the same site but vary in apparent height between 140 – 180 pm. This behavior differs from the behavior of Au and Pt adatoms, which preferentially adsorb in the V$_{O2c}$ of the TiO$_2$(110) surface [35]. Our data suggest no preferential interaction between Rh and V$_{O2c}$ as the density of the visible V$_{O2c}$ is identical before and after Rh deposition. This result agrees with our DFT calculations, where the best adsorption of Rh$_1$ is located at the hollow between the bridging oxygen row and the Ti row.

Figure 6d shows XPS spectra and the corresponding STM images of the Rh 3d peak of the as-deposited Rh at 100 K and after subsequent annealing to 873 K. Between 100 K and 150 K the Rh 3d peak shifts to higher binding energy. This core level shift could be related to a final state effect linked to the small size of the clusters, which appear in STM at 150 K. A similar effect can be recognised when comparing Au$_1$ and Au$_3$ to Au nanoparticles on the reduced

TiO$_2$(110) surface. The Au 4f binding energies of the Au$_1$ and Au$_3$ are similar to those at a higher coverage of Au. This effect occurs from a cancellation of initial and final state effects for Au$_1$ and Au$_3$. By increasing the coverage, the binding energy first decreases and finally increases [71]. Between 150 K and 250 K the peak stays the same but when heating the sample to temperatures above 300 K the peak shifts gradually to lower binding energy. This is related to the formation of bigger clusters with increasing temperature. These experimental observations, where Rh is seen to cluster after heating just to 150 K (Fig. 6d) agree with our previous calculated diffusion barriers and preference to sintering.

### 4. Discusssion and conclusions

The presence of localized charges on surfaces in form of small polarons, unavoidably impacts adsorption and reaction processes and surface dynamics. Previous studies have shown that adsorbed CO exhibit different coupling regime (from attractive to repulsive) depending on the position and density of small polarons, revealing a polaron-mediated correlation between CO adsorption energy and reduction state of the sample [13]. Substantial polaron-charge transfer has been found in O$_2$ adsorption, leading to the formation of superoxo and peroxo species [14]. Here, we have shown that single atom adsorption is also strongly coupled with polaron-charge transfer effect, which affects the adsorption energy and, importantly, the oxidation state of the metal atoms. For example, a Pt$^{2-}$ binds stronger than a Pt$^-$ at the V$_O$ site by 0.39 eV. This also implies that the reduction level and associated polaron density does affect the stability of the adsorbed metal atoms, and therefore its surface dynamics, but also its reactivity (i.e. charge transfer between the single-atom catalysts and adsorbed molecules).

For our model system, we found that adatoms of Pt$_1$ and Au$_1$ have low diffusion barriers and preferably adsorb in V$_{O2c}$. At higher dosing amounts, clusters can be observed. Rather than forming polarons, the V$_{O2c}$ excess electrons are transferred to the metal atoms, altering their electronic structure by filling their valence states and result in modifying their relative stabilities. Rh$_1$ adatoms, however show no preference for the V$_{O2c}$ defect, leaving the excess charge to form polarons and instead adsorb at hollow sites, which is in agreement with our experimental STM observations where they quickly sinter into clusters with increasing temperature at low coverage.

The interaction between adsorbates and polarons is further complicated by presence of additional species which could alter the surface charge balance. In this respect, it is also important to note that the TiO$_2$(110) surface is fully oxidized in a realistic condition. Therefore, V$_O$ is repaired by H$_2$O or O$_2$ molecules, leading to hydroxylated TiO$_2$(110) or oxidized TiO$_2$(110) surfaces. The oxidized TiO$_2$(110) surface was suggested to bind Au$_1$ stronger than the V$_O$ site on the reduced TiO$_2$(110) surface [53]. Contrary to the Au case, only Pt clusters were observed on the oxidized TiO$_2$(110) surface.

Nevertheless, the presence of small polarons is important even for a realistic TiO$_2$(110) surface, because repair of an oxygen vacancy by H$_2$O results in two electron polarons arising from two surface OH groups. Clearly, the modelling of such complicated multi-polaron configurational space (which might also involve the simultaneous formation of electron and hole polarons) is clearly unfeasible via conventional DFT. In order to efficiently explore the energy landscape, novel automated machine-learning based methods must be implemented and employed to determine the most favorable (structural and electronic) configurations and the most likely dynamical paths [72].

**Acknowledgments**

G.S.P., P.S., L.H., M.M. and M.A. acknowledge funding the European Research Council (ERC) under the European Union's Horizon 2020 research and innovation programme (Grant Agreement No. 864628). Z.J. was supported by the Austrian Science Fund (FWF, Y847-N20, START Prize). C.F., G.S.P., M.R. and M.M. were supported by the Austrian Science Fund (FWF) project SFB TACO (F81).

**References**

1.  Yang X, Wang A, Qiao B, Li JUN (2013) Single-Atom Catalysts : A New Frontier. 46:1740-1748

2.  Liu J (2017) Catalysis by Supported Single Metal Atoms. ACS Catal 7:34–59. https://doi.org/10.1021/acscatal.6b01534

3.  Zhang H, Liu G, Shi L, Ye J (2018) Single-Atom Catalysts: Emerging Multifunctional Materials in Heterogeneous Catalysis. Adv Energy Mater 8:1–24. https://doi.org/10.1002/aenm.201701343

4.  Qiao B, Wang A, Yang X, et al (2011) Single-atom catalysis of CO oxidation using Pt1/FeOx. Nat Chem 3:634–641. https://doi.org/10.1038/nchem.1095

5.  Jones J, Xiong H, DeLaRiva AT, et al (2016) Thermally stable single-atom platinum-on-ceria catalysts via atom trapping. Science 353:150–154. https://doi.org/10.1126/science.aaf8800

6.  Liang S, Hao C, Shi Y (2015) The Power of Single-Atom Catalysis. ChemCatChem 7:2559–2567. https://doi.org/10.1002/cctc.201500363

7.  Gates BC, Flytzani-Stephanopoulos M, DIxon DA, Katz A (2017) Atomically dispersed supported metal catalysts: Perspectives and suggestions for future research. Catal Sci Technol 7:4259–4275. https://doi.org/10.1039/c7cy00881c

8.  Chen F, Jiang X, Zhang L, et al (2018) Single-atom catalysis: Bridging the homo- and heterogeneous catalysis. Cuihua Xuebao/Chinese J Catal 39:893–898. https://doi.org/10.1016/S1872-2067(18)63047-5

9.  Lang R, Xi W, Liu JC, et al (2019) Non defect-stabilized thermally stable single-atom catalyst. Nat Commun 10:1–10. https://doi.org/10.1038/s41467-018-08136-3

10. Zhang J, Wu X, Cheong WC, et al (2018) Cation vacancy stabilization of single-atomic-site Pt 1 /Ni(OH) x catalyst for diboration of alkynes and alkenes. Nat Commun 9:1–8. https://doi.org/10.1038/s41467-018-03380-z

11. Wan J, Chen W, Jia C, et al (2018) Defect Effects on TiO2 Nanosheets: Stabilizing Single Atomic Site Au and Promoting Catalytic Properties. Adv Mater 30 https://doi.org/10.1002/adma.201705369

12. Franchini C, Reticcioli M, Setvin M, Diebold U (2021) Polarons in materials. Nat Rev Mater 6:560–586. https://doi.org/10.1038/s41578-021-00289-w

13. Reticcioli M, Sokolović I, Schmid M, et al (2019) Interplay between Adsorbates and Polarons: CO on Rutile TiO2 (110). Phys Rev Lett 122:1–6. https://doi.org/10.1103/PhysRevLett.122.016805

14. Sokolović I, Reticcioli M, Čalkovský M, et al (2020) Resolving the adsorption of molecular O2 on the rutile TiO2(110) surface by noncontact atomic force microscopy. Proc Natl Acad Sci U S A 117:14827–14837. https://doi.org/10.1073/pnas.1922452117

15. Cao Y, Yu M, Qi S, et al (2017) Scenarios of polaron-involved molecular adsorption on reduced TiO2(110) surfaces. Sci Rep 7:1–7. https://doi.org/10.1038/s41598-017-06557-6

16. Papageorgiou AC, Beglitis NS, Pang CL, et al (2010) Electron traps and their effect on the surface chemistry of TiO 2(110). Proc Natl Acad Sci U S A 107:2391–2396. https://doi.org/10.1073/pnas.0911349107

17. Gono P, Wiktor J, Ambrosio F, Pasquarello A (2018) Surface Polarons Reducing Overpotentials in the Oxygen Evolution Reaction. ACS Catal 8:5847–5851. https://doi.org/10.1021/acscatal.8b01120

18. Lv CQ, Liu JH, Guo Y, et al (2016) DFT + U investigation on the adsorption and initial decomposition of methylamine by a Pt single-atom catalyst supported on rutile (110) TiO 2. Appl Surf Sci 389:411–418. https://doi.org/10.1016/j.apsusc.2016.07.111

19. Wang J, Zhang W, Zhu W, et al (2020) Rutile TiO2 supported single atom Au catalyst: A facile approach to enhance methanol dehydrogenation. Mol Catal 482:110670. https://doi.org/10.1016/j.mcat.2019.110670

20. Fung V, Hu G, Tao F, Jiang D en (2019) Methane Chemisorption on Oxide-Supported Pt Single Atom. ChemPhysChem 20:2217–2220. https://doi.org/10.1002/cphc.201900497

21. Ammal SC, Heyden A (2017) Titania-Supported Single-Atom Platinum Catalyst for Water-Gas Shift Reaction. Chemie-Ingenieur-Technik 89:1343–1349. https://doi.org/10.1002/cite.201700046

22. Chang TY, Tanaka Y, Ishikawa R, et al (2014) Direct imaging of pt single atoms adsorbed on TiO2 (110) Surfaces. Nano Lett 14:134–138. https://doi.org/10.1021/nl403520c

23. Tang Y, Asokan C, Xu M, et al (2019) Rh single atoms on TiO2 dynamically respond to reaction conditions by adapting their site. Nat Commun 10:1–10. https://doi.org/10.1038/s41467-019-12461-6

24. Reticcioli M, Setvin M, Schmid M, et al (2018) Formation and dynamics of small polarons on the rutile TiO2 (110) surface. Phys Rev B 98:45306. https://doi.org/10.1103/PhysRevB.98.045306

25. Setvin M, Franchini C, Hao X, et al (2014) Direct view at excess electrons in TiO2 rutile and anatase. Phys Rev Lett 113:1–5. https://doi.org/10.1103/PhysRevLett.113.086402

26. Janotti A, Franchini C, Varley JB, et al (2013) Dual behavior of excess electrons in rutile TiO2. Phys Status Solidi - Rapid Res Lett 7:199–203. https://doi.org/10.1002/pssr.201206464

27. Nørskov JK, Bligaard T, Rossmeisl J, Christensen CH (2009) Towards the computational design of solid catalysts. Nat Chem 1:37–46. https://doi.org/10.1038/nchem.121

28. Corma A (2016) Heterogeneous catalysis: Understanding for designing, and designing for applications. Angew Chemie - Int Ed 55:6112–6113. https://doi.org/10.1002/anie.201601231

29. Zhao ZJ, Chiu CC, Gong J (2015) Molecular understandings on the activation of light hydrocarbons over heterogeneous catalysts. Chem Sci 6:4403–4425. https://doi.org/10.1039/c5sc01227a

30. Li L, Chang X, Lin X, et al (2020) Theoretical insights into single-atom catalysts. Chem Soc Rev 49:8156–8178. https://doi.org/10.1039/d0cs00795a

31. Chrétien S, Metiu H (2007) Density functional study of the interaction between small Au clusters, Aun (n=1–7) and the rutile TiO2 surface. II. Adsorption on a partially reduced surface. J Chem Phys 127:244708. https://doi.org/10.1063/1.2806802

32. Pillay D, Hwang GS (2005) Growth and structure of small gold particles on rutile TiO2(110). Phys Rev B - Condens Matter Mater Phys 72:1–6. https://doi.org/10.1103/PhysRevB.72.205422

33. Pillay D, Wang Y, Hwang GS (2004) A comparative theoretical study of Au, Ag and Cu adsorption on TiO 2 (110) rutile surfaces. Korean J Chem Eng 21:537–547. https://doi.org/10.1007/BF02705445

34. Okazaki K, Morikawa Y, Tanaka S, et al (2004) Electronic structures of Au on TiO2(110) by first-principles calculations. Phys Rev B - Condens Matter Mater Phys 69:1–8. https://doi.org/10.1103/PhysRevB.69.235404

35. Mellor A, Humphrey D, Yim CM, et al (2017) Direct visualization of au atoms bound to TiO2(110) O-vacancies. J Phys Chem C 121:24721–24725. https://doi.org/10.1021/acs.jpcc.7b09608

36. Kresse G, Furthmüller J (1996) Efficiency of ab-initio total energy calculations for metals and semiconductors using a plane-wave basis set. Comput Mater Sci 6:15–50. https://doi.org/10.1016/0927-0256(96)00008-0

37. Vargas-Hernández RA (2020) Bayesian Optimization for Calibrating and Selecting Hybrid-Density Functional Models. J Phys Chem A 124:4053–4061. https://doi.org/10.1021/acs.jpca.0c01375

38. Blöchl PE (1994) Projector augmented-wave method. Phys Rev B 50:17953–17979. https://doi.org/10.1103/PhysRevB.50.17953

39. Joubert D (1999) From ultrasoft pseudopotentials to the projector augmented-wave method. Phys Rev B - Condens Matter Mater Phys 59:1758–1775. https://doi.org/10.1103/PhysRevB.59.1758

40. Dion M, Rydberg H, Schröder E, et al (2004) Van der Waals density functional for general geometries. Phys Rev Lett 92:22–25. https://doi.org/10.1103/PhysRevLett.92.246401

41. Klimeš J, Bowler DR, Michaelides A (2010) Chemical accuracy for the van der Waals density functional. J Phys Condens Matter 22. https://doi.org/10.1088/0953-8984/22/2/022201

42. Klime J, Bowler DR, Michaelides A (2011) Van der Waals density functionals applied to solids. Phys Rev B - Condens Matter Mater Phys 83:1–13. https://doi.org/10.1103/PhysRevB.83.195131

43. Giustino F (2017) Electron-phonon interactions from first principles. Rev Mod Phys 89:1–63. https://doi.org/10.1103/RevModPhys.89.015003

44. Maxisch T, Zhou F, Ceder G (2006) Ab initio study of the migration of small polarons in olivine Lix FePO4 and their association with lithium ions and vacancies. Phys Rev B - Condens Matter Mater Phys 73:1–6. https://doi.org/10.1103/PhysRevB.73.104301

45. Nolan M, Watson GW (2006) Hole localization in Al doped silica: A DFT+U description. J Chem Phys 125. https://doi.org/10.1063/1.2354468

46. Wang Z, Brock C, Matt A, Bevan KH (2017) Implications of the DFT+U method on polaron properties in energy materials. Phys Rev B 96:1–13. https://doi.org/10.1103/PhysRevB.96.125150

47. Deskins NA, Rousseau R, Dupuis M (2011) Distribution of Ti3+ surface sites in reduced TiO2. J Phys Chem C 115:7562–7572. https://doi.org/10.1021/jp2001139

48. Di Valentin C, Pacchioni G, Selloni A (2006) Electronic structure of defect states in hydroxylated and reduced rutile TiO2(110) surfaces. Phys Rev Lett 97. https://doi.org/10.1103/PhysRevLett.97.166803

49. Henkelman G, Uberuaga BP, Jónsson H (2000) Climbing image nudged elastic band method for finding saddle points and minimum energy paths. J Chem Phys 113:9901–9904. https://doi.org/10.1063/1.1329672

50. Henkelman G, Jónsson H (2000) Improved tangent estimate in the nudged elastic band method for finding minimum energy paths and saddle points. J Chem Phys 113:9978–9985. https://doi.org/10.1063/1.1323224

51. Fernández-Torre D, Yurtsever A, Onoda J, et al (2015) Pt atoms adsorbed on TiO2(110)-(1×1) studied with noncontact atomic force microscopy and first-principles simulations. Phys Rev B - Condens Matter Mater Phys 91:1–8. https://doi.org/10.1103/PhysRevB.91.075401

52. Wang X, Zhang L, Bu Y, Sun W (2021) Interplay between invasive single atom Pt and native oxygen vacancy in rutile TiO2(110) surface: A theoretical study. Nano Res 12:4–11. https://doi.org/10.1007/s12274-021-3542-5

53. Matthey D, Wang JG, Wendt S, et al (2007) Enhanced bonding of gold nanoparticles on oxidized TiO2(110). Science (80- ) 315:1692–1696. https://doi.org/10.1126/science.1135752

54. Giordano L, Pacchioni G, Bredow T, Sanz JF (2001) Cu, Ag, and Au atoms adsorbed on TiO2(1 1 0):

Cluster and periodic calculations. Surf Sci 471:21–31. https://doi.org/10.1016/S0039-6028(00)00879-7

55. Allen JP, Watson GW (2014) Occupation matrix control of d- and f-electron localisations using DFT + U. Phys Chem Chem Phys 16:21016–21031. https://doi.org/10.1039/c4cp01083c

56. Huber F, Giessibl FJ (2017) Low noise current preamplifier for qPlus sensor deflection signal detection in atomic force microscopy at room and low temperatures. Rev Sci Instrum 88. https://doi.org/10.1063/1.4993737

57. Choi JIJ, Mayr-Schmölzer W, Mittendorfer F, et al (2014) The growth of ultra-thin zirconia films onPd3Zr(0 0 0 1). J Phys Condens Matter 26. https://doi.org/10.1088/0953-8984/26/22/225003

58. Lun Pang C, Lindsay R, Thornton G (2008) Chemical reactions on rutile TiO2(110). Chem Soc Rev 37:2328–2353. https://doi.org/10.1039/b719085a

59. Diebold U (2003) Structure and properties of TiO2 surfaces: A brief review. Appl Phys A Mater Sci Process 76:681–687. https://doi.org/10.1007/s00339-002-2004-5

60. Diebold U (2003) The surface science of titanium dioxide. Surf Sci Rep 48:53–229

61. Deák P, Aradi B, Frauenheim T (2012) Quantitative theory of the oxygen vacancy and carrier self-trapping in bulk TiO 2. Phys Rev B - Condens Matter Mater Phys 86:1–8. https://doi.org/10.1103/PhysRevB.86.195206

62. Moses PG, Janotti A, Franchini C, et al (2016) Donor defects and small polarons on the TiO2(110) surface. J Appl Phys 119. https://doi.org/10.1063/1.4948239

63. Valentin C Di, Pacchioni G, Selloni A (2009) Reduced and n-Type Doped TiO 2 : Nature of Ti 3 + Species. J Phys Chem C 113:20543–20552

64. Diebold U, Lehman J, Mahmoud T, et al (1998) Intrinsic defects on a TiO2(110)(1 × 1) surface and their reaction with oxygen: A scanning tunneling microscopy study. Surf Sci 411:137–153. https://doi.org/10.1016/S0039-6028(98)00356-2

65. Brookes IM, Muryn CA, Thornton G (2001) Imaging water dissociation on TiO2(110). Phys Rev Lett 87:266103-1-266103–4. https://doi.org/10.1103/PhysRevLett.87.266103

66. Schaub R, Thostrup P, Lopez N, et al (2001) Oxygen vacancies as active sites for water dissociation on rutile TiO2(110). Phys Rev Lett 87:266104-1-266104–4. https://doi.org/10.1103/PhysRevLett.87.266104

67. Wendt S, Schaub R, Matthiesen J, et al (2005) Oxygen vacancies on TiO2(1 1 0) and their interaction with H2O and O2: A combined high-resolution STM and DFT study. Surf Sci 598:226–245. https://doi.org/10.1016/j.susc.2005.08.041

68. Du Y, Deskins NA, Zhang Z, et al (2009) Imaging consecutive steps of O 2 reaction with hydroxylated TiO 2(110): Identification of HO 2 and terminal OH intermediates. J Phys Chem C 113:666–671. https://doi.org/10.1021/jp807030n

69. Li L, Li W, Zhu C, Mao LF (2021) A DFT+U study about agglomeration of Au atoms on reduced surface of rutile TiO2 (110). Mater Chem Phys 271:124944. https://doi.org/10.1016/j.matchemphys.2021.124944

70. Hugenschmidt MB, Gamble L, Campbell CT (1994) The interaction of H2O with a TiO2(110) surface. Surf Sci 302:329–340. https://doi.org/10.1016/0039-6028(94)90837-0

71. Mellor A, Wilson A, Pang CL, et al (2020) Photoemission core level binding energies from multiple sized nanoparticles on the same support: TiO2(110)/Au. J Chem Phys 152. https://doi.org/10.1063/1.5135760

72. Birschitzky VC, Ellinger F, Diebold U, et al (2022) Machine Learning for Exploring Small Polaron Configurational Space. https://doi.org/10.48550/arXiv.2202.01042